# SUPERCONDUCTIVITY

# Resolving thermoelectric "paradox" in superconductors

Connor D. Shelly,[1]* Ekaterina A. Matrozova,[2] Victor T. Petrashov[1]†

For almost a century, thermoelectricity in superconductors has been one of the most intriguing topics in physics. During its early stages in the 1920s, the mere existence of thermoelectric effects in superconductors was questioned. In 1944, it was demonstrated that the effects may occur in inhomogeneous superconductors. Theoretical breakthrough followed in the 1970s, when the generation of a measurable thermoelectric magnetic flux in superconducting loops was predicted; however, a major crisis developed when experiments showed puzzling discrepancies with the theory. Moreover, different experiments were inconsistent with each other. This led to a stalemate in bringing theory and experiment into agreement. With this work, we resolve this stalemate, thus solving this long-standing "paradox," and open prospects for exploration of novel thermoelectric phenomena predicted recently.

## INTRODUCTION

A temperature gradient applied to a superconductor, S, will yield an electric current $\vec{j}_q = -\eta_q \nabla T$, where $\eta_q$ is the thermoelectric coefficient. It is carried by quasiparticles, the unpaired "normal" charge carriers that exist at finite temperatures in any superconductor and behave similarly to conduction electrons in a normal metal. However, in contrast to a normal metal, the total current in the bulk of a homogeneous superconductor vanishes because the thermoelectric current, $\vec{j}_q$, is canceled by a superconducting current $\vec{j}_s$, ensuring that the total current in the bulk is zero: $\vec{j} = \vec{j}_s + \vec{j}_q = 0$ to obey the Meissner effect (1, 2). What renders the thermoelectric supercurrent measurable is its relation to the gradient of the macroscopic phase, $\theta(\vec{r})$, of the superconducting condensate wave function $\psi_s(\vec{r}) = (n_s/2)^{1/2} \exp(i\theta(\vec{r}))$, where $n_s$ is the concentration of "superconducting" electrons (3). Combining $\vec{j}_s = (e\hbar n_s/2m)\nabla\theta(\vec{r})$ and $\vec{j}_s = -\vec{j}_q = \eta_q \nabla T$, we find that the application of a temperature gradient to the superconductor creates a phase difference $\Delta\theta = (2m\eta_q/e\hbar n_s)\Delta T$ across the superconductor. The phase difference is analogous to the difference in electrochemical potential across the ends of a normal metal placed in a temperature gradient (4). When the ends of the superconductor, S, are connected by a dissimilar superconductor, S′, making a closed loop, as in Fig. 1A, a circulating supercurrent, $I_{cs}$, flowing within penetration depth, $\Lambda$, is induced, generating a thermoelectric magnetic field, $B_{Th}$, and a corresponding thermoelectric magnetic flux $\Phi_{Th}$.

Practical measurements of thermoelectric magnetic flux became feasible with the emergence of extremely sensitive magnetometers based on superconducting quantum interference devices (SQUIDs) (5). A SQUID contains a superconducting loop interrupted by "weak links" (5), making it sensitive to the magnetic flux $\Phi_1$ threading the loop. When the loop is coupled to the bimetallic superconducting loop via mutual inductance, M, the flux $\Phi_1$ becomes dependent on the supercurrent $I_2$ circulating in the bimetallic loop: $\Phi_1 = \Phi_{1e} - MI_2$, where $\Phi_{1e}$ is the flux through the interferometer loop area generated by sources other than the bimetallic loop. The circulating supercurrent $I_{cs}$ and the associated thermoelectric flux $\Phi_{Th}$ can be measured when one of the contacts of the bimetallic loop is heated, thus creating a temperature gradient. The first thermoelectric flux measurement (6) was in a reasonable agreement with the existing theory. However, further experiments (4, 7, 8) showed temperature-dependent magnetic fluxes up to five orders of magnitude larger than predicted by the theory (3) with unexpected dependence on the temperature. From an experimental viewpoint, the genuine thermoelectric effect could be masked by concomitant effects that may occur in the presence of a background field due to the temperature dependence of the penetration depth, $\Lambda$, of the superconductor (9). To minimize spurious $\Lambda$ effects, an ingenious experiment was undertaken using a bimetallic superconducting torus (4) to minimize contributions of the background magnetic field. Nevertheless, an unexpectedly large temperature-dependent flux was still observed. Such a large thermoelectric flux would imply "giant" values for the superconducting component of the thermoelectric current and similar values for the quasiparticle component, $\vec{j}_q$; yet, the experiments that followed revealed a paradoxical mismatch between the values of $\vec{j}_s$ and $\vec{j}_q$. The measured quasiparticle current (10) was orders of magnitude smaller than the supercurrent and agreed well with the theory (11). No plausible explanations for the observations and no experimentally supported analysis have been offered to date, and the paradox remained unresolved [see Galperin et al. (12) and Gurevich et al. (13) and references therein].

Here, we resolve this paradox. We explain the reason why the masking $\Lambda$ effects were not excluded completely in previous experiments and present experimental results of the separation of genuine thermoelectric flux from the masking effects. Theoretically, we deduce a new formula connecting the thermoelectric magnetic flux and the current circulating in the loop in the presence of a temperature gradient. For calculations, we use a method suggested recently (13), taking into account the energy balance in the system. Our proof-of-principle experiments agree well with the new theory.

In the presence of background external field, B, the current circulating in the bimetallic loop, $I_2$, is present even in the absence of the temperature gradient. It keeps the number, k, of flux quanta trapped in the bimetallic loop constant, in accordance with the quantization rule for the flux, $\Phi_2$, through the bimetallic loop: $\Phi_2 = \Phi_{2e} - L_2I_2 = k\Phi_0$, where $\Phi_{2e} = A_2B$ is the flux generated by external field, B, through the bimetallic loop area, $A_2$; $L_2$ is the self-inductance of the loop; and $k = 0, \pm1, \pm2$ and so on. As a result, the total flux through the interferometer loop is $\Phi_1 = \Phi_{1e} + (M/L_2)(A_2B + k\Phi_0)$. Heating of one of the contacts of the bimetallic loop not only creates a temperature gradient but also increases the average temperature of the loop. Although the temperature gradient induces a thermoelectric flux $\Phi_{Th}$, the rise in the temperature results in an increase of the penetration depth $\Lambda \rightarrow \Lambda + \delta\Lambda$ (Fig. 1A) with the corresponding changes of the effective area of the loop $\delta A_2 = (dA_2/d\Lambda)\delta\Lambda$ and inductances $\delta L_2 = (dL_2/d\Lambda)\delta\Lambda$ and $\delta M = (dM/d\Lambda)\delta\Lambda$. Taking into account that usually the changes are relatively small, $\delta A_2 << A_2$, $\delta L_2 << L_2$, and $\delta M << M$, the flux increment induced by heating is given as $\delta\Phi_1 = \delta(MA_2/L_2)B + \delta(M/L_2)k\Phi_0 + (M/L_2)\Phi_{Th}$.

[1]Department of Physics, Royal Holloway, University of London, Egham, Surrey TW20 0EX, UK. [2]Laboratory of Cryogenic Nanoelectronics, Nizhny Novgorod State Technical University, Nizhny Novgorod 603950, Russia.
*Present address: National Physical Laboratory, Teddington TW11 0LW, UK.
†Corresponding author. E-mail: v.petrashov@rhul.ac.uk

The signal measured by the quantum interferometer is a periodic function of $\Phi_1$ with the period equal to the flux quantum $\Phi_0 = h/2e$. In practice, the interferometer signal oscillations are measured as a function of magnetic field, $B$, with the period and phase of the oscillations depending on $M$, $L_2$, $A_2$, and the $k$ number. The flux increment, $\delta\Phi_1$, results in the phase shifts of the oscillations. They include the thermoelectric flux contribution (the third term) and contributions due to the changes in the penetration depth, the $\Lambda$ effect (the first and second terms). The first $\Lambda$ effect phase shift corresponds to the changes in the oscillations' period and depends on the ambient magnetic field. The second shift is $k$-dependent. The $\Lambda$ effect contributions may exceed the contribution of thermoelectric flux by several orders of magnitude (4, 8, 9). Although the first phase shift can be calculated using precise enough measurements of the period of oscillations at different temperature gradient values, the second shift is practically indistinguishable from the thermoflux contribution when the control of the number of flux quanta, $k$, trapped in the bimetallic loops is lacking. Moreover, fluctuations in the $k$ values from one experiment to another may lead to scatter in the measured values. The previous experiments used macroscopic loops of several millimeters in diameter that corresponded to about $10^6$ trapped flux quanta due to the geomagnetic field. To precisely determine and control the $k$ value in such large loops as well as measure the period of oscillations, one had to control the absolute magnitude of magnetic field with a precision of about $10^{-9}$ T, an extremely difficult task that was never done. Consequently, separation of the genuine thermoelectric flux has not been completed, leaving the puzzling discrepancy between the theory and the experiment unexplained (1, 12, 13). The conceptual and technological advance reported here is based on highly sensitive hybrid quantum interferometers (14, 15) coupled to the bimetallic loops fabricated using modern nanolithography with areas up to five orders of magnitude smaller than those measured previously. This allowed both precise control of the number of trapped flux quanta and measurements of the changes in the oscillations' period under the temperature gradient. The thermoflux is separated using characteristic phase shifts in the quantum interference oscillations that are independent of the applied magnetic field and $k$ numbers. The measured thermoelectric flux changes its sign upon reversal of the direction of the temperature gradient, a distinctive feature of a genuine thermoelectric effect.

## RESULTS

### Theory of thermoelectric magnetic flux

The thermoelectric magnetic flux, $\Phi_{Th}$, and the circulating current, $I_{cs}$, are connected by the formula

$$\Phi_{Th} = L_2 I_{cs} \quad (1)$$

We consider a superconducting thermocouple consisting of a superconductor, $S$, and a superconductor, $S'$, with a much larger gap and, hence, negligible quasiparticle current. This situation corresponds with our experiments; however, it is simple to generalize it to an arbitrary superconductor $S'$.

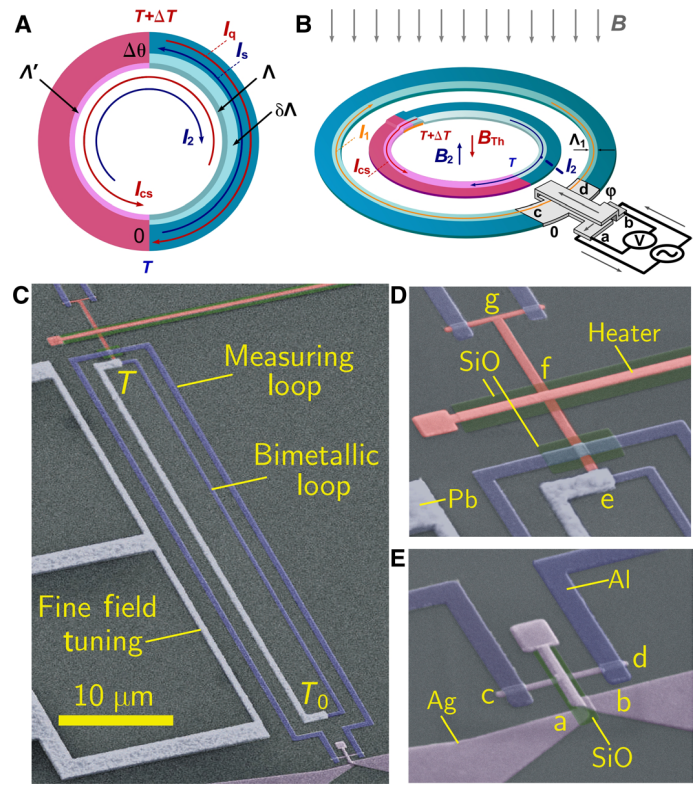

**Fig. 1. Bimetallic superconducting loop and experimental setup for thermoelectric flux measurements.** (**A**) Bimetallic superconducting loop in the temperature gradient. Thermoelectric quasiparticle current, $I_q$, in the bulk of the superconductor with a smaller gap is opposed by counterflowing supercurrent $I_s$ with superconducting phase difference, $\Delta\theta$, created and circulating current $I_{cs}$ induced within the penetration depth, $\Lambda$; with an increase in average temperature, the penetration depth acquires an increment, $\delta\Lambda$, that results in an increase in the effective area of the loop; the screening current $I_2$ keeps the total magnetic flux through the bimetallic loop constant. (**B**) Diagram of experimental setup. A bimetallic loop made of different superconductors is placed within a loop of a hybrid quantum interferometer a-b-c-d. The total flux through the bimetallic loop, $\Phi_2$, is created by the external magnetic field $B$ and the fields $B_{Th}$ and $B_2$ induced by thermoelectric circulating current $I_{cs}$ and screening current $I_2$. The interferometer measures the superconducting phase difference $\phi$ between c and d that is proportional to the total magnetic flux through the interferometer loop. This includes the flux through the bimetallic loop and externally induced flux that is partially screened by the current $I_1$. (**C**) False-colored scanning electron micrograph of a bimetallic loop coupled to a heater and hybrid quantum interference device (HyQUID). (**D**) A fluxless heater f generating a temperature gradient in the bimetallic loop by local spot heating of the contact e. The temperature $T_1$ at e is measured by superconductor/normal/superconductor (SNS) thermometer g. (**E**) The HyQUID measuring a superconducting phase difference between c and d with folded normal wires insulated by a spacer [see (B) for details].

By the current conservation, the circulating current is independent of the coordinate, $x$, along the bimetallic loop, $I_q(x) + I_s(x) = I_{cs}$ = const, where $I_q$ and $I_s$ are the bulk thermoelectric quasiparticle and superconducting currents, respectively. The theory of thermoelectric quasiparticle current, $I_q$, was developed by Galperin et al. (3), Guénault and Webster (*16*), and Van Harlingen *et al.* (*4*); however, calculation of the superconducting circulating current, $I_{cs}$, which generates the thermoelectric magnetic flux, was not attempted. Here, we proceed with the calculation using minimization of the total energy of the superconducting loop, $W$, with respect to $I_{cs}$, as suggested by Gurevich *et al.* (*13*); see the Supplementary Materials for details.

The total energy consists of the energy of the magnetic field created by the circulating current, $W_m = L_2 I_{cs}^2/2$, and the kinetic energy of superconducting electrons, $W_k = L_k I_s^2/2$, where $L_k$ is the kinetic inductance. We find that the circulating current is defined by the quasiparticle thermoelectric current, $I_q(0)$, and the kinetic inductance, $L_k(0)$, of the weaker superconductor at the hot contact: $I_{cs} = I_q(0) L_k(0)/(L_2 + L_k(0))$.

Next, we calculate the quasiparticle current at the hot contact, $I_q(0) = -\eta_q(0)(dT_q/dx)s$, where $s$ is the cross-sectional area of the superconductor. We use the heat equation that relates the rate at which the quasiparticles accumulate the energy from the heat flow, $\dot{Q}_q = -\kappa_q s(dT_q/dx)$, and the rate of quasiparticle energy transfer to the "local" phonons in the wire, $\dot{Q}_{qp} = \Sigma(T_q^n - T_p^n)sdx$; $T_p$ is the phonon temperature, $\kappa_q$ is the quasiparticle thermal conductivity, and $\Sigma$ is a material electron-phonon interaction parameter (*17*). We obtain $I_q(0) = (\eta_q(0)/\kappa_q(0))\langle \Sigma(T_q^n - T_p^n)\rangle sl_0$, where $\kappa_q(0)$ is the thermal conductivity at the hot contact and the brackets ⟨ … ⟩ mean an average over the length $l_0$ of the weaker superconductor, with $n = 5$ or 6 depending on the relationship between the electron mean free path, $l$, and the phonon wavelength, $\lambda_p$ (*18*, *19*). Using connections of the coefficients $\eta_q(0)$, $\kappa_q(0)$ and $\Sigma$ in the superconducting state to their values in the normal state, $\eta_N(0)$, $\kappa_N(0)$ (*3*, *20*) and $\Sigma_N$ (*21*), and assuming that the normal state quasiparticle thermal conductivity $\kappa_N(0)$ obeys the Wiedemann-Franz law $\kappa_N(0) = \mathcal{L}\sigma_N T_q(0)$, where $\mathcal{L} = 2.4 \times 10^{-8}$ V$^2$/K$^2$ is the Lorenz number and $\sigma_N$ is conductivity of the smaller gap superconductor in the normal state, we arrive at the following formula for the thermoelectric flux

$$\Phi_{Th} = -\alpha_N \Sigma_N \langle G(T_q)(T_q^n - T_p^n)\rangle L_{eff} \mathcal{V}/\mathcal{L} \quad (2)$$

where $\alpha_N = \eta_N(0)/\sigma_N T_q(0)$ (V/K$^2$) is the gradient in the dependence of thermopower on temperature that is constant in the investigated temperature range [see, for example, Mamin *et al.* (*10*)], $L_{eff} = L_2 L_k(0)/(L_2 + L_k(0))$ is the effective inductance, $\mathcal{V}$ is the volume of the small-gap superconductor, and the function $G(T_q)$ describes the drop in electron-phonon interaction upon the temperature decrease below the critical temperature. The value of the gap at the hot contact is the smallest within the bimetallic loop, making the contact essentially a weak link. The kinetic inductance of such a link is given by $L_k = \Phi_0/2\pi I_c$, where $I_c$ is the critical current (*22*). At temperatures close to critical, the kinetic inductance diverges as the critical current goes to zero. The value of the circulating current approaches that of the quasiparticle current, $I_{cs} \to I_q(0)$. According to Eq. 2, the thermoflux originates from the temperature difference between the heated electron system and the phonon system. It is a continuous function at the critical temperature and above it. We started with finite supercurrent $I_s$; however, no singularities in the thermoflux are expected at the critical temperature. When the bulk supercurrent fades, the value of circulating current $I_{cs}$ continuously approaches the value of quasiparticle current, $I_q$. This is a result of current conservation. Simultaneously, kinetic inductance diverges at the critical temperature and the effective inductance approaches the value of geometric inductance, $L_2$.

The formula gives several conditions to maximize the thermoflux in the superconducting state. Besides an obvious condition of the large thermopower, the formula requires large effective inductance and strong electron-phonon interaction as well as coupling of the "local" phonon system to the environment of the bimetallic loop to ensure difference in the electron and phonon temperatures. The formula is valid when one of the superconductors enters the normal state. In this case, by the current conservation, the circulating current is equal to the thermoelectric current in the normal wire, $I_{cs} = I_N(0)$; in addition, in this case, $G(T_q) = 1$ and $L_{eff} = L_2$.

## Experiments

Our experimental setup is shown in Fig. 1. The bimetallic loop is placed inside the measuring loop of a quantum magnetometer (Fig. 1, B and C). The characteristic magnetic field $B = \Phi_0/A_2 \approx 10^{-5}$ T was high enough to allow precise control of the $k$ number.

We used a magnetometer based on a HyQUID (*14*, *23*, *24*). Similar to a standard SQUID, the magnetometer consisted of a superconducting loop interrupted by a weak link. The weak link consisted of a normal conductor connected to the superconductor at c and d (Fig. 1, B to E). The electrical resistance $R$ between points a and b is an oscillating function of the total magnetic flux $\Phi_1$ threading the interferometer loop

$$R = R_0 - r\cos(2\pi\Phi_1/\Phi_0) \quad (3)$$

where $R_0$ is independent of magnetic field and $r$ is the amplitude of oscillations. To determine the thermoelectric flux, we measured several resistance oscillations as a function of applied magnetic field. The values $B_{n,k}$ of the field at the extrema of the resistance correspond to $\Phi_1 = n\Phi_0$, with $n = (2m + 1)/2$ for maxima and $n = m$ for minima; $m = 0, \pm 1, \pm 2, \pm 3$ and so on. In the absence of the temperature gradient, the measured values $B_{n,k}$ can be found using the formula

$$B_{n,k}A + (M/L_2)k\Phi_0 = n\Phi_0 \quad (4)$$

where $A = A_1 - (M/L_2)A_2$ is the effective area of the measuring interferometer loop and $A_1$ is its actual area (see Materials and Methods for details).

Examples of oscillations as a function of applied magnetic field are shown in Fig. 2. Figure 2A shows the oscillations in the absence of the temperature gradient at different $k$ numbers. The $k$ numbers were manipulated by thermocycling of the bimetallic loop between the normal state and the superconducting state in the precalculated applied magnetic field. When the $k$ number was changed to $k + \Delta k$, the positions of the extrema were shifted by $\Delta B$, which was constant for all maxima and equal to $\Delta B = (M/L_2)(\Phi_0/A)\Delta k$ with $(M/L_2) = 0.2$, which is in agreement with our design value. When the temperature gradient is established, the total flux $\Phi_1 = n\Phi_0$ corresponding to a particular extremum of the resistance is intact, whereas the measured positions of the extrema are shifted to the new values $B'_{n,k}$ satisfying the equation $B'_{n,k}A' + (M'/L'_2)(k\Phi_0 - \Phi_{Th}) = n\Phi_0$, where $A' = A + \delta A$ and $(M'/L'_2) = (M/L_2) + \delta(M/L_2)$ are the values modified by the Λ effect. We used the difference $\Delta B_{n,k} = B_{n,k} - B'_{n,k}$, the phase shift in oscillations, to determine the thermoelectric flux

$$\frac{\Delta B_{n,k}}{B_0} = an - bk - c \quad (5)$$

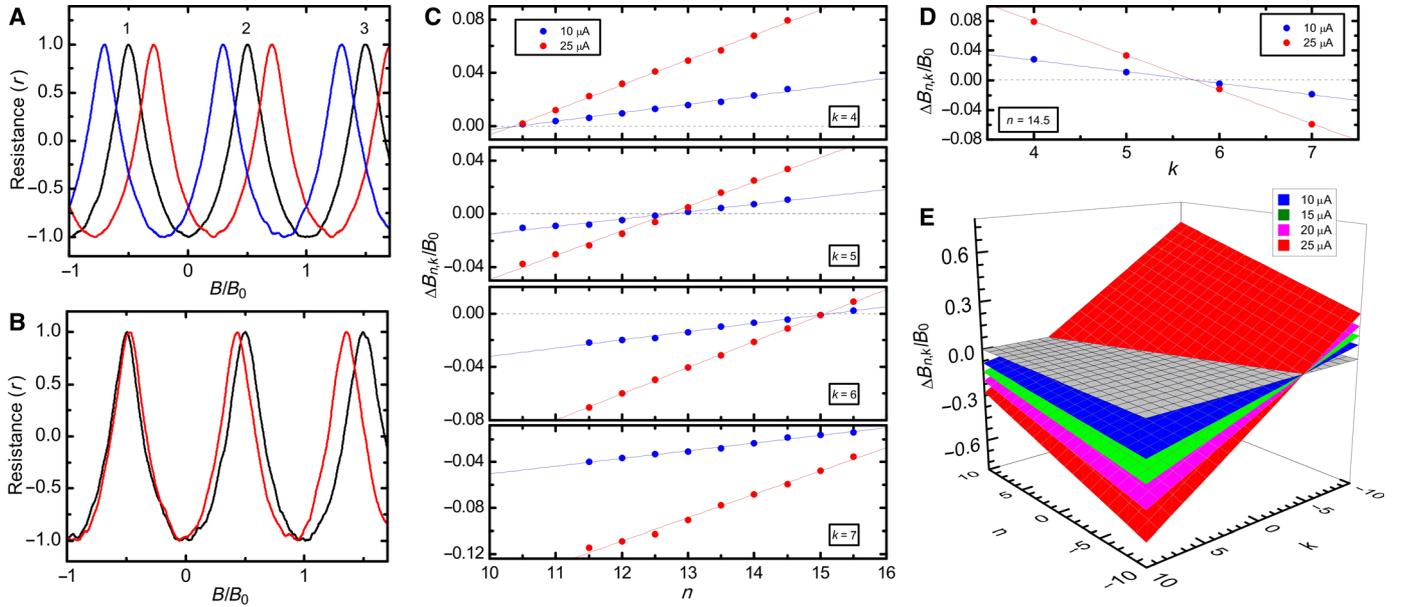

**Fig. 2. Phase shifts in oscillations at different $k$ numbers and heater currents.** (**A**) Oscillations at different $k$ numbers of magnetic flux quanta trapped in the bimetallic loop with the heater current OFF. Left maximum corresponds to $k = 4$; middle, $k = 5$; right, $k = 6$; 1, $n = 8.5$; 2, $n = 9.5$; 3, $n = 10.5$. A change in $k$ number introduces constant phase shift, leaving the period intact. (**B**) Oscillations with the heater current OFF (red line) and with the heater current ON (black line). (**C**) The shifts $\Delta B_{n,k}$ [as in (B)] of the values $B_{n,k}$ with the heater current ON versus the total flux $\Phi_1 = n\Phi_0$ at different fixed $k$ numbers of the trapped flux quanta. (**D**) The shifts $\Delta B_{n,k}$ versus $k$ number at a fixed total flux. (**E**) Three-dimensional (3D) representation of the shifts $\Delta B_{n,k}$ versus $k$ number and the total flux $\Phi_1 = n\Phi_0$ through the measurement loop. The planes correspond to different fixed heater currents $I_h = 10$, 15, 20, and 25 μA.

The coefficients $a = \delta A/A$ and $b = \delta(M/L_2)$ describe the phase shifts due to the Λ effect, and the phase shift $c$ is associated with the thermoelectric flux, $\Phi_{Th}$, and is independent of $n$, $k$ and magnetic field; $B_0 = \Phi_0/A$ is the period of oscillations. To create a temperature gradient, we used a heater made of a normal metallic silver wire (Fig. 1, C and D) connected to one of the contacts of the bimetallic loop. Figure 2B shows the oscillations at a fixed $k$ number at different heater currents. The heater produced two effects described in Eq. 5: (i) the phase shift associated with the thermoelectric flux, $\Phi_{Th}$, and (ii) the change in the period of oscillations due to the Λ effect.

The dependencies $\Delta B_{n,k}/B_0$ versus $n$ and $k$ at fixed heater currents are shown in Fig. 2 (C and D). They are linear, in agreement with Eq. 5. The slopes give the values $a$ and $b$.

Equation 5 at a constant heater current is an equation for a plane in 3D space with $x$, $y$, and $z$ coordinates corresponding to $n$, $k$ and $\Delta B_{n,k}/B_0$. The planes plotted using measurement results are shown in Fig. 2E. They allow visualization of the Λ effect contribution $(an - bk)$ at different $k$ and $n$ values at a glance. The Λ effect is minimal along the line where the planes intersect. The $n$ and $k$ numbers along the line satisfy the relation $an \approx bk$; hence, keeping ratio $k/n$ constant and changing $p = \text{Integer}(bk/an)$, we move along the line. For this particular sample, $b/a \approx 5/2$ and $k/n = 2/5$. To decrease error in the value of $c$, we averaged measurements at different combinations of $n$ and $k$. The measured values of $c$ as a function of $p$ at several values $n$ and $k$ are shown in Fig. 3A. As expected, the value of $c$ that is directly related to thermoelectric flux was independent of $k$ and $n$ and depended on the heater current. Here, we emphasize that to obtain the value of the thermoflux, one has to know the absolute values of $n$ and $k$. Thermocycling adds integers to the $k$ numbers: $k = k_0 + m$, where $m = \pm 1, \pm 2$, etc. and $k_0 = \text{Integer}(A_2 B_r/\Phi_0)$ is the number of flux quanta trapped because of the residual magnetic field, $B_r$, in the solenoid. We have measured $B_r$ using sensitive Hall probes and found it to be close to the geomagnetic field $B_g = 41$ μT in our case. With $\Phi_0/A_2 = 10$ μT, we find $k_0 = 4$. In previous experiments, this number was up to five orders of magnitude larger, leading to a much larger Λ effect (8, 9). Next, we have undertaken an experiment to verify that the measured thermoflux shows the signature of the genuine thermoflux with a change of the sign when the temperature gradient is reversed. We have nanofabricated a mirror-reflected bimetallic loop with the aluminum part replacing lead and vice versa. This is equivalent to a change in the direction of the temperature gradient and circulation of the thermoelectric current generating thermoelectric flux. The values of $c$ versus $p$ for the two samples are plotted in Fig. 3A. The sign of $c$ changes when the temperature gradient is reversed, as expected for the genuine thermoflux.

## DISCUSSION

To compare the measured values of the thermoflux with theory, we take into account the following aspects of our experiments. The thermal resistance of the wire/substrate interface (the Kapitza resistance) for our aluminum films was negligible because the thickness, $h = 80$ nm, was much less than the phonon wavelength, $\lambda_p \approx hv_s/k_B T \geq 200$ nm at $T \leq 1.2$ K, where $v_s \approx 5 \times 10^3$ ms$^{-1}$ is the sound velocity. The phonon temperature in such films is close to the substrate temperature, $T_p \approx T_{sub}$ (21), and the phonon term, $T_p^n$, in the difference, $T_q^n - T_p^n$, as seen in Eq. 2, can be neglected because the substrate temperature is estimated to be lower than the quasiparticle temperature $T_{sub} < T_q$; therefore, $T_q^n - T_p^n = T_q^n(1 - (T_{sub}/T_q)^n) \approx T_q^n$ (17). Second, to calculate the effective inductance $L_{eff}$, we consider the aluminum at the hot contact as essentially a weak link within the bimetallic loop. This is justified because the gap in aluminum is much smaller than that in lead. The kinetic inductance of such a link is given by $L_k = \Phi_0/2\pi I_c$, where $I_c$ is the critical current (22). The absolute value of thermoelectric flux at temperatures close to critical is $\Phi_{Th}(T_c) \approx \alpha_N \Sigma_N \langle T_c^n \rangle L_{eff} \mathcal{V}/\mathcal{L}$, with $L_{eff}(T_c) \approx L_2$, because the kinetic inductance at the hot end acquires large values $L_k(0) \gg L_2$ when the critical current approaches zero at $T \approx T_c$. At temperatures below critical, the value of the thermoflux can be written as $\Phi_{Th}(T_q) = \Phi_{Th}(T_c)\langle G(T_q/T_c) \cdot (T_q/T_c)^n\rangle(L_{eff}/L_2)$ with scaling factors describing the decrease in quasiparticle concentration, the electron-phonon interaction, and the decrease in the kinetic inductance (see the Supplementary Materials for details).

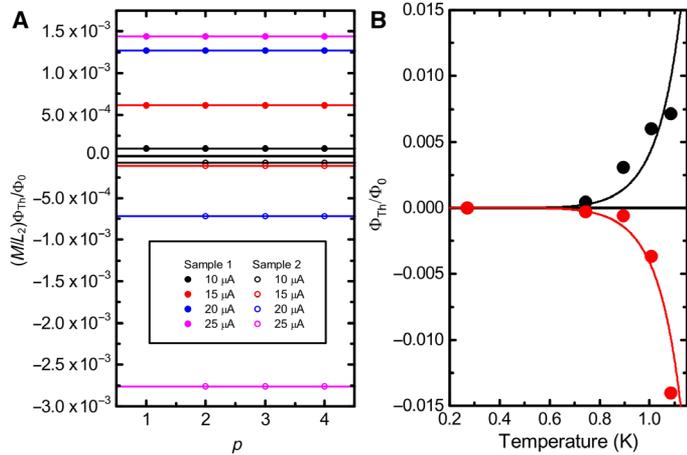

**Fig. 3. Thermoelectric phase shifts and magnetic flux at different heater currents and hot contact temperatures.** (**A**) The values of $\Delta B_{n,k}$ versus $p$ at different $k$ and $n$ values close to the line $\Delta B_{n,k} = 0$ in Fig. 2E with $\Lambda$ effect excluded. (**B**) The measured thermoelectric flux values at different hot spot temperatures for two samples with opposite direction of the temperature gradient (symbols). Curves are theory predictions according to Eq. 2.

The experimentally measured values fitted with the theoretical curves are shown in Fig. 3B. We used the calculated value $L_2 \approx 10^{-10}$ H for the geometrical inductance of our bimetallic loops, the cross section $s = 6 \times 10^{-14}$ m$^2$ and the length $l_0 = 5 \times 10^{-5}$ m, and critical temperature $T_c \approx 1.2$ K. Using a typical value of the electron mean free path $l \approx 50$ nm and the wavelength of phonons in our aluminum films calculated above, one would expect the films to belong to the strong disorder limit $l \ll \lambda_p$ with the electron-phonon interaction described by $n = 6$ (18). However, as shown in the Supplementary Materials, the results of fit with $n = 5$ do not differ much from that with $n = 6$ in the investigated temperature range. Both of the fits give the value of the product of the thermopower, $\alpha_N$, and the electron-phonon interaction parameter in the normal state, $\Sigma_N$: $\alpha_N \Sigma_N \approx 10^3$ VW/K$^{n+2}$m$^3$. Using the sample-dependent values of $\alpha_N$ in a realistic range, $2 \times 10^{-8}$ to $6 \times 10^{-7}$ V/K$^2$, we obtain the sample-dependent values of $\Sigma_N$ in the range $2 \times 10^9$ to $5 \times 10^{10}$ W/m$^3$K$^n$, which is in agreement with previous measurements (25). To fit the dependence of the thermoflux on temperature, we have taken the dependence of critical current at the hot contact in the form $I_c = I_0(1 - T_q(0)/T_c)$ (26) with a fitting parameter value of $I_0 = 30$ μA, which is realistic for our wires (see the Supplementary Materials for more comments on the temperature dependence).

In conclusion, we have developed a method allowing the separation of the genuine thermoelectric flux from spurious $\Lambda$ effects. Theoretically, we deduce the formula for the circulating current by minimizing the total energy of the system. We solve the heat equation for the loop and calculate the thermoelectric flux using the relation of the circulating current to the heat flow and the energy transfer rate from quasiparticles to phonons. The measured value of the thermoelectric flux agrees well with the theory. The technological and theoretical advance reported here opens a new avenue for exploration of thermal phenomena in superconducting devices (27–30), which is critical for energy management in nanoscale structures.

## MATERIALS AND METHODS

### Magnetometry with HyQUID

The HyQUID is especially suitable for this experiment. Its fabrication technology and composition are compatible with the other elements, which allows the whole structure to be placed on the same chip. This avoids the use of macroscopic leads to connect the bimetallic loop and the magnetometer, thus preventing stray signal pickup and the introduction of parasitic inductance (5). The device is designed to reduce coupling of the flux-sensitive loop with the measuring electronics: the normal segment of the flux-sensitive loop makes contact with the measuring wire a-b at a single point. As a result, the current noise is not transferred to the interferometer loop from the measuring circuit. To reduce inductive coupling of the flux-sensitive loop with the measuring circuit, the a-b wire was folded (see Fig. 1, B, C, and E, for details). Furthermore, the device was designed to make the voltage between the N/S interfaces negligible, thus reducing the Josephson radiation and extra noises that are typical in standard SQUID-based magnetometers. The interferometer proved to be a sensitive probe for minute supercurrents (23).

The total magnetic flux $\Phi_1$ in Eq. 3 threading the interferometer loop was $\Phi_1 = BA_1 - L_1I_1 - MI_2$, where $B$ is the applied magnetic field; $A_1$ and $L_1$ are the area and inductance of the interferometer loop, respectively; and $I_1$ is the Josephson screening current in the interferometer loop $I_1 = I_{c1} \sin(2\pi\Phi_1/\Phi_0)$. To avoid hysteresis, the HyQUID was designed so that the screening factor was less than unity: $L_1I_{c1} < 1$. The measured values $B_{n,k}$ in this regime were independent of the screening current because it completely vanishes at the extrema: $I_1 \propto \sin(2\pi n) = 0$. Calculating the current $I_2$ using quantization of the flux through the bimetallic loop $\Phi_2 = B_{n,k}A_2 - L_2I_2 = k\Phi_0$, we arrived at Eq. 5 for the measured values $B_{n,k}$.

The bimetallic loop was made of lead and aluminum with critical temperatures 7.2 and 1.2 K, respectively. Thus, the concentration of quasiparticles in lead was negligible in the investigated temperature range 0.25 to 1.1 K. The structure consisted of five layers. The first layer was silver; the second, spacers made of silicon monoxide; the third, upper silver layer; the fourth, aluminum; and the fifth, lead. The deposition was performed in medium vacuum of $10^{-6}$ Torr. To provide a clean interface for good electrical contact between the layers and good thermal contact with the substrate, we used in situ argon plasma etch. The thickness of silver, aluminum, silicon monoxide, and lead layers was 50, 80, 30, and 150 nm, respectively. The areas of measurement and bimetallic loops were $A_1 = 380$ μm$^2$ and $A_2 = 170$ μm$^2$, respectively. To apply magnetic field in a wide range, we used a superconducting solenoid. For fine field tuning, a wire made of lead was fabricated on the same chip (Fig. 1C). The measurements of the resistance $R$ were made using a standard low-frequency lock-in technique. To provide optimal operation of the magnetometer, the length of the c-d wire, $L_{cd}$, was made smaller than the electron phase breaking length, $L_\phi = \sqrt{D\tau_\phi}$, where $D$ is the diffusion coefficient and $\tau_\phi$ is the electron phase breaking time; however, $L_{cd}$ was large enough to suppress the Josephson current so that $L_1 I_1 < \Phi_0$ to avoid hysteresis (23, 24). These requirements defined the value of the resistance of the $L_{cd}$ wire, $R_{cd}$. The resistivity of silver was $\rho = 4 \times 10^{-8}\,\Omega \cdot m$, which gave the typical values of resistance $R_{cd} \approx 10\,\Omega$ for a given material and wire dimensions. The resistance of the $L_{ab}$ wire, $R_0$, in our interferometer was close to $R_{cd}$. The value of $r/R_0$ was in the range 2 to 15%, depending on the dimensions of the wires, in accordance with theory and previous experiments (14, 24, 31).

**Thermometry**
The temperature gradient was created by a heater made of a normal metallic silver wire (Fig. 1, C and D). The heater was made bifilar to minimize magnetic flux generated by the heating current. The residual flux induced by the heater was canceled by averaging the measurements with opposite heater currents. The heater was connected to one of the contacts of the bimetallic loop by a silver wire. To measure the temperature at the hot end, $T_q(0)$, we used the superconducting/normal/superconducting junction (32). The thermometer was placed away from the hot contact at point g to avoid coupling of the thermometer wires to the bimetallic and interferometer loops (Fig. 1, C and D). To make the temperatures at g and e (Fig. 1D) as close to each other as possible, the heater was thermally coupled to the thermometer and to the loop in a similar way. Points g and e were connected to the heater at point f by normal silver wires of the same length and cross-sectional area. Both wires had similar thermal boundary conditions at the substrate and the superconductors; hence, the temperature distribution in the wires obeyed similar heat equations. To calibrate the thermometer, we used the transition of aluminum in the bimetallic loop to the normal state that was measured using the dependence of penetration depth, $\Lambda$, on temperature (33); for details, see the Supplementary Materials. The substrate temperature $T_0$ close to the cold end was measured using the sensitivity of the amplitude of oscillations to the temperature; an increase in $T_0$ was negligible at the heater currents investigated.

**SUPPLEMENTARY MATERIALS**

Supplementary material for this article is available at http://advances.sciencemag.org/cgi/content/full/2/2/e1501250/DC1
Calculating thermoelectric flux
Thermometry
Fig. S1. Conversion of currents at the interface between two superconductors, S and S′, with different energy gaps, $\Delta$ and $\Delta'$.
Fig. S2. Calculated dependence of the thermoelectric flux scaling factors on the quasiparticle temperature normalized to critical temperature, $T_c$.
Fig. S3. Calculated temperature dependence of thermoelectric flux, $\Phi_{th}$, normalized to its value, $\Phi_{th}(T_c)$, at critical temperature with different values of critical current, $I_0$, and electron-phonon parameter, $n$.
Fig. S4. Dependence of critical current in the SNS thermometer on the length of the normal element at base temperature $T = 245$ mK.
Fig. S5. Differential resistance versus bias current curves for an SNS thermometer at different bath temperatures.
Fig. S6. Differential resistance versus bias current curves for an SNS thermometer at different heater currents.
Fig. S7. Temperature calibration curve.
Fig. S8. Determination of the heater current corresponding to the onset of superconductivity in the bimetallic loop.
Reference (34)

**REFERENCES AND NOTES**

1. V. L. Ginzburg, Nobel Lecture: On superconductivity and superfluidity (what I have and have not managed to do.) as well as on the "physical minimum" at the beginning of the XXI century. *Rev. Mod. Phys.* **76**, 981–998 (2004).
2. V. L. Ginzburg, On thermoelectric phenomena in superconductors. *J. Phys. USSR* **8**, 148 (1944).
3. Y. M. Galperin, V. L. Gurevich, V. I. Kozub, Thermoelectric effects in superconductors. *Zh. Eksp. Teor. Fiz.* **66**, 1387–1397 (1974).
4. D. J. Van Harlingen, D. F. Heidel, J. C. Garland Experimental study of thermoelectricity in superconducting indium. *Phys. Rev. B* **21**, 1842–1857 (1980).
5. J. Clarke, A. I. Braginski, *The SQUID Handbook: Vol. 1. Fundamentals and Technology of SQUIDs and SQUID Systems* (Wiley-VCH, Weinheim, 2004).
6. N. V. Zavaritskii, Observation of superconducting current excited in a superconductor by heat flow. *Zh. Eksp. Teor. Fiz. Pis'ma Red.* **19**, 205–208 (1974).
7. C. M. Falco, Thermally induced magnetic flux in a superconducting ring. *Solid State Commun.* **19**, 623–625 (1976).
8. D. J. Van Harlingen, J. C. Garland, Thermoelectric transport effect in superconducting indium. *Solid State Commun.* **25**, 419–422 (1978).
9. C. M. Pegrum, A. M. Guénault, Thermoelectric flux effects in superconducting bimetallic loops. *Phys. Lett. A* **59**, 393–395 (1976).
10. H. J. Mamin, J. Clarke, D. J. Van Harlingen, Charge imbalance induced by a temperature gradient in superconducting aluminum. *Phys. Rev. B* **29**, 3881–3890 (1984).
11. S. N. Artemenko, A. F. Volkov, The thermoelectric field in superconductors. *Zh. Eksp. Teor. Fiz.* **70**, 1051–1060 (1976).
12. Y. M. Galperin, V. L. Gurevich, V. I. Kozub, A. L. Shelankov, Theory of thermoelectric phenomena in superconductors. *Phys. Rev. B* **65**, 064531 (2002).
13. V. L. Gurevich, V. I. Kozub, A. L. Shelankov, Thermoelectric effects in superconducting nanostructures. *Eur. Phys. J. B* **51**, 285–292 (2006).
14. V. T. Petrashov, V. N. Antonov, P. Delsing, T. Claeson, Phase controlled conductance of mesoscopic structures with superconducting "mirrors". *Phys. Rev. Lett.* **74**, 5268–5271 (1995).
15. F. Giazotto, J. T. Peltonen, M. Meschke, J. P. Pekola, Superconducting quantum interference proximity transistor. *Nat. Phys.* **6**, 254–259 (2010).
16. A. M. Guénault, K. A. Webster, Thermally generated magnetic flux in superconducting loops. *J. Phys. Colloques* **39**, 539–540 (1978).
17. F. C. Wellstood, C. Urbina, J. Clarke, Hot-electron effects in metals. *Phys. Rev. B* **49**, 5942–5955 (1994).
18. J. T. Karvonen, L. J. Taskinen, I. J. Maasilta, Observation of disorder-induced weakening of electron-phonon interaction in thin noble-metal films. *Phys. Rev. B* **72**, 012302 (2005).
19. L. J. Taskinen, I. J. Maasilta, Improving the performance of hot-electron bolometers and solid state coolers with disordered alloys. *Appl. Phys. Lett.* **89**, 143511 (2006).
20. J. Bardeen, G. Rickayzen, L. Tewordt, Theory of the thermal conductivity of superconductors. *Phys. Rev.* **113**, 982–994 (1959).
21. A. V. Timofeev, C. Pascual García, N. B. Kopnin, A. M. Savin, M. Meschke, F. Giazotto, J. P. Pekola, Recombination-limited energy relaxation in a Bardeen-Cooper-Schrieffer superconductor. *Phys. Rev. Lett.* **102**, 017003 (2009).
22. K. K. Likharev, *Dynamics of Josephson Junctions and Circuits* (Gordon and Breach Science Publishers, New York, 1986).
23. V. T. Petrashov, K. G. Chua, K. M. Marshall, R. Sh. Shaikhaidarov, J. T. Nicholls, Andreev probe of persistent current states in superconducting quantum circuits. *Phys. Rev. Lett.* **95**, 147001 (2005).
24. V. T. Petrashov, R. Sh. Shaikhaidarov, I. A. Sosnin, P. Delsing, T. Claeson, A. Volkov, Phase-periodic proximity-effect compensation in symmetric normal/superconducting mesoscopic structures. *Phys. Rev. B* **58**, 15088–15093 (1998).
25. J. P. Kauppinen, J. P. Pekola, Electron-phonon heat transport in arrays of Al islands with submicrometer-sized tunnel junctions. *Phys. Rev. B* **54**, R8353–R8356 (1996).
26. M. Tinkham, *Introduction to Superconductivity* (McGraw-Hill, New York, ed. 2, 1996).
27. M. S. Kalenkov, A. D. Zaikin, L. S. Kuzmin, Theory of a large thermoelectric effect in superconductors doped with magnetic impurities. *Phys. Rev. Lett.* **109**, 147004 (2012).
28. P. Machon, M. Eschrig, W. Belzig, Nonlocal thermoelectric effects and nonlocal Onsager relations in a three-terminal proximity-coupled superconductor-ferromagnet device. *Phys. Rev. Lett.* **110**, 047002 (2013).
29. P. Machon, M. Eschrig, W. Belzig, Giant thermoelectric effects in a proximity-coupled superconductor-ferromagnet device. *New J. Phys.* **16**, 073002 (2014).
30. A. Ozaeta, P. Virtanen, F. S. Bergeret, T. T. Heikkilä, Predicted very large thermoelectric effect in ferromagnet-superconductor junctions in the presence of a spin-splitting magnetic field. *Phys. Rev. Lett.* **112**, 057001 (2014).
31. Y. V. Nazarov, T. H. Stoof, Diffusive conductors as Andreev interferometers. *Phys. Rev. Lett.* **76**, 823–826 (1996).
32. F. Giazotto, T. T. Heikkilä, A. Luukanen, A. M. Savin, J. P. Pekola, Opportunities for mesoscopics thermometry and refrigeration: Physics and applications. *Rev. Mod. Phys.* **78**, 217 (2006).

33. L. Hao, J. C. Macfarlane, J. C. Gallop, S. K. H. Lam. Direct measurement of penetration length in ultrathin and/or mesoscopic superconducting structures. *J. Appl. Phys.* **99**, 123916 (2006).
34. A. A. Abrikosov, in *Fundamentals of the Theory of Metals* (North-Holland, Amsterdam, 1988), pp. 465–470.

**Acknowledgments:** We gratefully acknowledge T. T. Heikkilä, M. Eschrig, A. F. Volkov, V. Zakosarenko, J. Goff, Yu. V. Nazarov, E. Il'ichev, A. D. Zaikin, and M. S. Kalenkov for fruitful discussions. **Funding:** The work was done using facilities of the Nanophysics and Nanotechnology Research Group at Royal Holloway, University of London, and was supported by British Engineering and Physical Sciences Research Council Grant EP/F016891/1. E.A.M. thanks L. S. Kuzmin for financial support within the Russian Ministry of Science Project no. 11.G34.31.0029. **Author contributions:** C.D.S. and E.A.M. designed and fabricated devices, performed the experiment, analyzed experimental data, and provided substantial contribution to the writing of the manuscript. V.T.P. conceived the project, contributed to the design of devices, performed the calculations, and wrote the manuscript. **Competing interests:** The authors declare that they have no competing interests. **Data and materials availability:** All data needed to evaluate the conclusions in the paper are present in the paper and/or the Supplementary Materials. Additional data related to this paper may be requested from the authors.

Submitted 10 September 2015
Accepted 24 December 2015
Published 26 February 2016
10.1126/sciadv.1501250

**Citation:** C. D. Shelly, E. A. Matrozova, V. T. Petrashov, Resolving thermoelectric "paradox" in superconductors. *Sci. Adv.* **2**, e1501250 (2016).